\documentstyle[preprint,revtex,eqsecnum]{aps}
\begin{document}
\draft
\begin{title}
Gott time machines cannot exist in an open\\ (2+1)-dimensional
universe with timelike total momentum \footnotemark
\footnotetext{Updated version, July 30, 1992: differs from
original by modifications in references.}
\end{title}
\author{Sean M. Carroll}
\begin{instit}
Harvard-Smithsonian Center for Astrophysics \\ 60 Garden Street,
Cambridge, Massachusetts 02138 U.S.A.
\end{instit}
\author{Edward Farhi and Alan H. Guth}
\begin{instit}
Center for Theoretical Physics, Laboratory for Nuclear Science, %
and Department of Physics,\\
Massachusetts Institute of Technology, %
Cambridge, Massachusetts 02139 U.S.A.
\end{instit}

\receipt{XXXXXXXXX}
\begin{abstract}
We prove the title.
\end{abstract}

\def\footnote#1#2{\footnotemark{#1} \footnotetext{#2}}

\narrowtext

\section{INTRODUCTION}

Absent some restriction on boundary conditions and energy
sources, it is possible for the spacetime metric of general
relativity to wreak havoc with our intuitive notions of ``going
forward in time.'' We can imagine metrics in which the worldline
of a test particle, locally restricted to the interior of its
forward light cone, can loop around to intersect itself --- a
closed timelike curve (CTC).  Indeed, it is easy to construct
solutions to Einstein's equations which exhibit such behavior
\cite{geroch,hawkingellis}.

Nonetheless, due to the causal paradoxes associated with such a
time machine, it is tempting to believe that CTC's exist only in
spacetimes that are in some way pathological.  That is, we would
expect that the laws of physics somehow act to prevent the
occurrence of CTC's in the real universe.  This expectation has
been dubbed the ``Chronology Protection Conjecture''
\cite{hawking}.

Scientific interest in CTC's was recently invigorated by Gott's
\cite{gott,allensimon} construction of an extraordinarily simple
solution to Einstein's equations that contains CTC's.  Gott's
solution represents two infinitely long parallel cosmic strings
moving past each other at high velocity.  The situation at early
times is portrayed in Fig.~1, which shows the two strings
approaching each other.  Each string is associated with a deficit
angle removed from the space, which we have oriented in a
direction opposite to that of the motion of the string.  Opposite
sides of the excluded wedges are identified at equal times.  Gott
found that, as the strings approach each other, it becomes
possible to traverse a closed timelike curve enclosing the
strings in the sense opposite to their motion.  Since many models
of the early universe suggest that it was populated by cosmic
strings moving at high velocities, the Gott solution raises the
possibility that CTC's could occur in the evolution of the real
world.

The Gott universe is topologically equivalent to Minkowski
space,\footnotemark \footnotetext{In discussing the topology of
the Gott universe, we are treating the strings as physical
objects with a small but nonzero thickness.  They are nonsingular
configurations, and are not excised from the spacetime.}  free of
singularities and event horizons, and does not violate the weak
energy condition.  Cutler \cite{cutler} has shown that the
spacetime contains regions free of CTC's, and indeed that there
are complete spatial hypersurfaces both to the future and the
past of the region containing CTC's.  Cutler constructed such a
hypersurface, but an even simpler example is provided by the
constant-time slice portrayed in Fig.~1; the past light cone of
any point on this surface extends through similar surfaces
arbitrarily far into the past, implying that no timelike curve
through such a point can be closed.  As the two strings approach
one another, each will ultimately collide with the trailing
deficit angle of the other, at which time this coordinate system
will fail; CTC's will then arise.

The Gott universe is known to contain CTC's of arbitrarily large
spatial extent \cite{ori,cutler,DJT}.  However, even if we
consider arbitrarily large CTC's to be unrealistic, this fact by
itself is no justification for rejecting the Gott time machine as
a physical possibility.  Since the universe does not consist of
an isolated Gott time machine, the interesting question is
whether a Gott time machine can arise within a universe populated
by other objects as well.  These other objects could affect the
boundary conditions at spatial infinity, so the properties of the
isolated solution at spatial infinity are not
decisive.\footnotemark \footnotetext{An analogy showing how one
might be misled by considering a boundary condition at spatial
infinity is the case of the 't~Hooft-Polyakov magnetic monopole.
A physicist studying the classical behavior of a grand unified
theory might choose to impose a boundary condition requiring the
Higgs field to approach a constant at infinity.  This boundary
condition would exclude a net magnetic charge, and hence the
monopole solution.  Nonetheless, monopole-antimonopole pair
configurations are consistent with the boundary condition, and
such configurations might evolve even if they are not included in
the initial configuration.  The monopoles arising in this way
would only locally resemble the ideal 't~Hooft-Polyakov solution,
yet they would exhibit the same physical characteristics.  Thus,
although the stringent boundary condition can exclude the
isolated monopole solution, it cannot prevent physical monopoles
from arising in the theory.} Since the Cauchy problem of general
relativity is formulated in terms of initial data on a complete
spacelike hypersurface, our goal is to isolate the
characteristics of those initial data that lead to CTC's.

We are thus led to study the conditions under which Gott time
machines can arise.  The issue of time machine construction in
3+1 dimensions has been addressed by Tipler \cite{tipler} and
Hawking \cite{hawking}, as we will further discuss in
\hbox{Sec.~III}.  In this paper, however, we address the question
in the context of (2+1)-dimensional gravity.  The
(2+1)-dimensional approach is based on the observation that a
spacetime populated solely by infinitely long parallel cosmic
strings is invariant with respect to boosts and translations
along the direction of the strings.  The dependence of the metric
on this direction is trivial and can be disregarded, in which
case the strings are equivalent to point masses in (2+1)
dimensions.  Each point particle is assigned a mass $M$
equivalent to the mass per unit length of the cosmic string.
This theory has been the object of extensive investigation
\cite{marderetal,DJTorig}. It has been found that the metric in
vacuum is necessarily flat, while in the presence of a single
particle with mass $M$, the external metric is that of Minkowski
space from which a wedge of angle $\alpha=8\pi GM$ has been
removed and opposite sides have been identified ($G$ is Newton's
constant).  Solutions with several static particles are easily
constructed by joining several one-particle solutions, in which
case the space has a net deficit angle given by the sum of the
deficit angles of the constituent particles. If the total deficit
angle exceeds $2\pi$ the spatial sections of the spacetime must
be closed, with topology $S^2$, and the total deficit angle is
necessarily exactly $4\pi$.  By joining appropriately boosted
single-particle spacetimes, the exact solutions with moving
particles \cite{DJTorig} and nontrivial decays and scatterings
\cite{CFG} may be constructed.

In their seminal paper on (2+1)-dimensional gravity, Deser,
Jackiw, and 't~Hooft \cite{DJTorig} noted that a spinning point
particle would give rise to CTC's.  They added, however, that
``such closed timelike contours are not possible in a space with
$n$ moving spinless particles, where angular momentum is purely
orbital.'' When stated without qualification, this sentence is
apparently contradicted by the existence of the Gott solution.
In this paper we find out what qualifications need to be added to
this sentence in order to validate it.  We will prove rigorously
that there exists a class of initial conditions for
(2+1)-dimensional universes from which a Gott time machine can
never evolve.

The energy and momentum of a collection of particles can be
conveniently characterized by the Lorentz transformation that a
vector undergoes upon being parallel transported around the
system \cite{DJTorig,wittenetal}.  The Lorentz transformation
belongs to the three-dimensional Lorentz group, SO(2,1).  Deser,
Jackiw, and 't~Hooft \cite{DJT} have shown that the group element
corresponding to the Gott time machine is boostlike ---
equivalent under similarity transformation to a pure boost.  Each
element of SO(2,1) can be identified with a 3-vector, which we
will refer to as the energy-momentum vector of the system. The
energy-momentum vector for a single particle is timelike, while
that of the Gott two-particle system is spacelike (tachyonic),
despite the fact that each particle is moving slower than $c$.
(The notion of a tachyonic momentum is used somewhat loosely in
this context; we will explain this more fully below.) This state
of affairs can arise because the momentum of a system of
particles is not the sum of the individual momenta.  Thus (2+1)
dimensional gravity offers the possibility of constructing
tachyonic momenta from ordinary, subluminal matter.

We are accustomed to thinking of tachyons as ``unphysical'', but
it does not follow as an immediate corollary that the Gott time
machine cannot be built.  The situation here is different from
special relativity--- in special relativity one can choose not to
introduce tachyons as fundamental particles, and the linear rule
for combining energy-momentum vectors then guarantees that no set
of particles will have a tachyonic momentum.  The corresponding
relation in (2+1)-dimensional gravity, however, is nonlinear,
allowing the construction of a tachyon from timelike particles.
The isolated Gott time machine, of course, can be excluded by
choosing to consider only universes for which the total momentum
is timelike.  However, if we want to consider the possibility of
a Gott time machine existing in conjunction with other matter,
then the question becomes more complicated.  We would like to
know whether a universe with a total momentum that is timelike
can contain a subsystem of particles for which the combined
momentum is spacelike.  Note that in special relativity,
spacelike momenta can be combined to form a timelike momentum.
If this can also happen in (2+1)-dimensional gravity, then it is
conceivable that a universe populated by slowly moving particles,
in which there is no hint of a time machine at early times, could
evolve by a series of decays and scatterings into a universe
containing a Gott time machine.

In an open universe, such is not the case --- we show in this
paper that no collection of particles with net timelike momentum
can contain a subset with spacelike momentum.  In our previous
paper \cite{CFG} we presented the physical reason behind this
result: there can never be enough mass in an open universe to
accelerate two particles to sufficiently high velocity.  In this
paper we provide the formal proof.  Specifically, we associate
with every collection of particles an element of the universal
covering group of SO(2,1), and show that an element corresponding
to an open universe with timelike momentum is never the product
of an element representing the Gott time machine and any number
of elements representing massive particles.  The proof can be
constructed by algebraic manipulation; however, an elegant
geometric demonstration is achieved by introducing an invariant
metric on the parameter space of the group, in which case the
group manifold becomes three-dimensional anti-de Sitter
space.\footnotemark \footnotetext{We are very grateful to Don
Page and Alex Lyons, who pointed out to us the relationship
between the (2+1)-dimensional Lorentz group and anti-de Sitter
space.}

In a closed universe, this result no longer holds --- we have
found that it is possible to construct a spacetime in which a
subset of particles with tachyonic momentum is created from
initially static conditions ({\it e.g.}, by the decay of massive
particles).  However, 't~Hooft \cite{thooft} has shown that
causal disaster is avoided, since the universe shrinks to zero
volume before any CTC's can arise.\footnotemark \footnotetext{In
a note added to our previous paper \cite{CFG} we erroneously
claimed that CTC's would arise.  We thank G. 't~Hooft for
informing us of our mistake.} He goes on to argue that this
phenomenon will always result from an attempt to build a time
machine in a closed universe. Therefore, neither closed nor open
universes can evolve Gott time machines from initial conditions
with slowly-moving particles. At the same time, the fact that a
pair of Gott particles can be constructed from static conditions
in a closed universe serves to emphasize the fact that a Gott
pair is not intrinsically unphysical.  It is therefore
interesting to understand exactly under what conditions a Gott
pair can and cannot be constructed.

\section{THEOREM AND PROOF}

\proclaim Theorem.  Consider an open (2+1)-dimensional universe
composed of point particles, each with a future-directed momentum
that is either timelike or lightlike. If the total momentum of
this universe is timelike, no subgroup of these particles can
have a spacelike momentum.

\noindent Since the Gott time machine consists of two particles
with net spacelike momentum, it follows immediately that such a
time machine can exist in an open universe only if the total
momentum of the universe is spacelike.

\subsection{Outline}

The outline of the proof is as follows.  We first review how to
characterize the energy and momentum of a gravitating particle by
an element of the Lorentz group SO(2,1).  The element for a
single particle in its rest frame is simply a rotation by its
deficit angle $\alpha$. The element for a moving particle is
obtained by a Lorentz transformation, and the element for a
collection of particles is the product of those associated with
its constituents. We then endow the parameter space of SO(2,1)
with an invariant metric, establishing the correspondence between
the group manifold and anti-de Sitter space.  We note that the
curves in SO(2,1) defined by $T(\lambda)=\exp(-i\lambda X)$,
where $X$ is a fixed element of the Lie algebra, are geodesics of
this metric passing through the identity.  A particle with a
future-directed timelike (lightlike) momentum is associated with
an element of SO(2,1) lying on a future-directed timelike
(lightlike) geodesic through the identity. The element associated
with a collection of particles corresponds to a nonspacelike
curve constructed from consecutive geodesic segments
corresponding to the individual particles.

An open timelike universe (i.e., an open universe with a timelike
total momentum) is associated (in its rest frame) with a rotation
matrix $R(\theta_0)$, such that $\theta_0\leq 2\pi$.  We would
like to show that there is no future-directed timelike or
lightlike curve from the element associated with the Gott time
machine to an element associated with an open timelike universe.
However, a slight complication arises due to the periodicity of
SO(2,1), in which a rotation by $\theta + 2 \pi$ is equivalent to
a rotation by $\theta$.  Starting from the Gott time machine, one
can easily find future-directed timelike curves that use this
periodicity to return to small values of $\theta$, terminating at
an open timelike universe.  The total mass of a (2+1)-dimensional
universe, however, is not periodic, and there is no way that
particles with a total deficit angle of $2.2\,\pi$ can be put
together to make an open universe.  This complication is
eliminated by replacing SO(2,1) with its universal covering
group, within which the rotation angle $\theta$ can take any
value from $- \infty$ to $\infty$ without identification.  In the
parameter space of the universal covering group, it can be shown
that open timelike universes cannot be reached from the Gott time
machine by future-directed timelike or lightlike curves.
Consequently, one cannot construct an open universe with timelike
total momentum that consists of a Gott time machine and any set
of point particles with future-directed timelike or lightlike
momenta.

We wish to comment that many of the tools we use are standard
results in the theory of Lie groups and symmetric spaces
\cite{helgason}. Any Lie algebra has a natural metric, the
Cartan-Killing form, given in components by
\begin{equation}
    g_{\mu\nu}=c^\lambda{}_{\mu\sigma} c^\sigma{}_{\nu\lambda} \ ,
  \eqnum{1}
\end{equation}
where the $c^{\alpha}{}_{\beta\gamma}$ are the structure
constants. For semisimple groups, this metric can be uniquely
extended to a left- and right-invariant metric on the entire
group manifold. The resulting space will be maximally symmetric,
and paths from the identity defined by $T(\lambda)=\exp(-i\lambda
X)$, where $X$ is an element of the Lie algebra, will be
geodesics of this metric.  However, it is straightforward for us
to derive these results for the case at hand, which we will do
for clarity.  (A similar discussion of the universal cover of
SO(2,1) in a different context can be found in
Ref.~\cite{balogetal}.  A more mathematical discussion can be
found in Ref.~\cite{pukansky}.)

\subsection{Energy and Momentum Conservation}

We begin by recalling the characterization of energy and momentum
in (2+1) dimensional gravity \cite{DJTorig,wittenetal}.  In any
number of dimensions, spacetime curvature can be characterized by
the Lorentz transformation (holonomy) that describes the result
of parallel transporting a vector around a closed loop.  In (2+1)
dimensions this technique is especially convenient, since a
closed curve unambiguously divides the space of worldlines into
those which pass through the loop and those which do not.
Further, since space is flat outside sources, any two loops
enclosing the same matter distribution will yield the same
transformation.

A vector parallel transported counterclockwise around a single
particle of deficit angle $\alpha$ is transformed by a
counterclockwise rotation matrix $R(\alpha)$. For a particle
moving with velocity $\vec v$, the appropriate transformation can
be obtained by boosting to the particle's rest frame, rotating,
and boosting back.  Thus, we associate with the moving particle a
matrix
\begin{equation}
    T=B(\vec \xi)R(\alpha)B^{-1}(\vec \xi) \ ,\eqnum{2}
\end{equation}
where $\vec \xi =\hat v \tanh^{-1}|\vec v|$ is the rapidity of
the particle and $B(\vec \xi)$ is a boost bringing the rest
vector to the velocity of the particle.  A loop around a system
of several particles can be deformed to a series of loops around
the individual particles.  Once an ordering is specified, the
transformation associated with a system is therefore given by the
product of the one-particle matrices:
\begin{equation}
    T_{tot}=T_N T_{N-1}\ldots T_1 \ .\eqnum{3}
\end{equation}
As a system of $N$ particles evolves via decay or scattering into
a system of $M$ particles, a loop around the system at one time
can be deformed to a loop around the system at a later time.
Since the deformation carries the loop only through regions of
flat spacetime, the resulting transformation matrix is not
changed. Therefore, conservation of energy and momentum is
expressed as the equality of Lorentz transformations at different
times:
\begin{equation}
    T_N T_{N-1}\ldots T_1=T^\prime_M T^\prime_{M-1}\ldots
     T^\prime_1 \ . \eqnum{4}
\end{equation}
This rule was used in Ref.~\cite{CFG} in constructing the
spacetime for one particle decaying into two.

We find it convenient to represent the transformations $T$ by $2
\times 2$ matrices.  The standard basis for the Lie algebra of
SO(2,1) consists of the rotation generator $J$ and two boost
generators $K_1$ and $K_2$, with the commutation relations
\begin{eqnarray}
     &&[K_1,K_2] = -iJ \nonumber \\
     &&[J,K_1] = i K_2 \nonumber \\
     &&[J,K_2] = - i K_1 \ .\eqnum{5}
\end{eqnarray}
Following the conventions used in Ref.~\cite{CFG}, we take $J =
\frac{1}{ 2} \sigma_3$ and $K_i = \frac{i}{2}
\sigma_i$, where the $\sigma$'s are the standard Pauli matrices.
When exponentiated, these $2 \times 2$ matrices generate the
group SU(1,1), which is a double cover of SO(2,1).

The generators $J$ and $K_i$ are components of an antisymmetric
Lorentz tensor $M_{\mu\nu}$, with $J = M_{12}$ and $K_i =
M_{i0}$.  Since we are working in three spacetime dimensions,
however, we can define a set of 3-vector group generators by
using the Levi-Civita tensor:
\begin{equation}
    {\cal J}_\mu = \frac{1}{2}\epsilon_\mu{}^{\lambda\sigma}
     M_{\lambda\sigma} \ ,\eqnum{6}
\end{equation}
where $\epsilon_{012} \equiv 1$, and indices are raised and
lowered by the Lorentz metric $\eta_{\mu \nu} \equiv \,{\rm
diag}\, [-1,1,1]$.  Explicitly, ${\cal J}_0 = J$, ${\cal J}_1 =
-K_2$, and ${\cal J}_2 = K_1$.

A rotation by angle $\alpha$ is given by
\begin{equation}
    R(\alpha)= e^{- i \alpha J} = \left\lgroup
     \matrix{e^{-i\alpha/2}&0\cr 0& e^{i\alpha/2}\cr}
     \right\rgroup \eqnum{7}
\end{equation}
and a boost is given by
\begin{equation}
    B(\vec \xi)=e^{- i \vec \xi \cdot \vec K} =
     \left\lgroup \matrix{\cosh\frac{\xi}{2}&e^{-i\phi}\sinh
     \frac{\xi}{2}\cr e^{i\phi}\sinh\frac{\xi}{2} &
     \cosh\frac{\xi}{2}\cr} \right\rgroup ,\eqnum{8}
\end{equation}
where $\xi=|\vec \xi |$ is the magnitude of the rapidity and
$\phi$ is its polar angle.  The matrix $T$ associated with a
single particle is found by evaluating Eq.~(2):
\begin{equation}
    T = \left\lgroup \matrix{ \sqrt{1 + p^2} e^{-i \alpha' /2} &
     i p e^{- i \phi}\cr - i p e^{i \phi} & \sqrt{1 + p^2} e^{i
     \alpha' /2}\cr} \right\rgroup ,\eqnum{9}
\end{equation}
where
\begin{equation}
    p \equiv \sinh \xi \sin \frac{\alpha}{2} \eqnum{10}
\end{equation}
is a measure of the momentum of the particle, and $\alpha^\prime$
is defined by
\begin{equation}
    \tan \frac{\alpha'}{2} = \cosh \xi \tan \frac{\alpha}{2} \
     .\eqnum{11}
\end{equation}
Note that $\alpha^\prime$ may be thought of as a boosted deficit
angle, when the excised wedge is taken to lie along the direction
of motion (as in Fig.~1).

\subsection{The Metric on SO(2,1)}

We now turn to the geometry of the parameter space of SO(2,1).
For convenience we continue to use the $2 \times 2$ matrix
representation. The space of $2 \times 2$ matrices is spanned by
the group generators ${\cal J}_\mu$ ($\mu=1,2,3$) and the
identity matrix, so an arbitrary $2 \times 2$ matrix can be
written as
\begin{equation}
    T = w - 2 i \chi^\mu {\cal J}_\mu =
     \left\lgroup \matrix{w - i t & y + i x\cr
     y - i x & w + i t\cr } \right\rgroup ,\eqnum{12}
\end{equation}
where $\chi^\mu \equiv (t,x,y)$, and $w$,$t$,$x$, and $y$ are
complex. SU(1,1), which is the double cover of SO(2,1), consists
of those matrices $T$ satisfying
\begin{equation}
    \det T=+1 \eqnum{13a}
\end{equation}
and
\begin{equation}
    T^\dagger \eta T= \eta \ ,\eqnum{13b}
\end{equation}
where
\begin{equation}
    \eta = \left\lgroup \matrix{1&0\cr 0&-1\cr} \right\rgroup .
     \eqnum{14}
\end{equation}
It is shown in the Appendix that these conditions are obeyed if
and only if $(t,x,y,w)$ are real numbers satisfying
\begin{equation}
    -t^2 + x^2 + y^2 - w^2 = -1 \ . \eqnum{15}
\end{equation}
The form of this equation suggests that we consider the resulting
three-dimensional space as embedded in a four-dimensional space
with metric
\begin{equation}
    ds^2=-dt^2 + dx^2 + dy^2 - dw^2 \ .\eqnum{16}
\end{equation}
It is natural to take the metric on the parameter space of
SU(1,1) to be the metric induced by this embedding.  Aficionados
of de~Sitter spaces will recognize this as three-dimensional
anti-de~Sitter space (see Ref.~\cite{hawkingellis}).  The group
SO(2,2) which leaves this metric invariant will map the
submanifold defined by Eq.~(15) into itself, so the embedded
three-manifold is maximally symmetric.  Furthermore, it will be
shown in the Appendix that the metric is group invariant--- that
is, either left- or right-translation by an element of SU(1,1) is
an isometry of the metric.

To put coordinates on SU(1,1), note that we can decompose any
element into a boost times a rotation:
\begin{eqnarray}
    T&=&B(\vec\zeta)R(\theta) \nonumber \\
  &=& \left\lgroup \matrix{e^{-i\theta/2}\cosh\frac{\zeta}{2}&
  e^{i(\theta/2-\delta)}\sinh\frac{\zeta}{2}\cr
  e^{i(-\theta/2+\delta)}\sinh\frac{\zeta}{2}&
  e^{i\theta/2}\cosh\frac{\zeta}{2}\cr} \right\rgroup ,\eqnum{17}
\end{eqnarray}
where $\zeta$ is the magnitude of the boost, $\delta$ is its
polar angle, and $\theta$ is the angle of rotation.  (Note that
this is a different parameterization than the variables
($\alpha$, $\xi$, $\phi$) used in Eqs.~(9-11).) Comparing (17) to
Eq.~(12) and defining $\psi=2\delta-\theta$, we obtain
\begin{eqnarray}
    t &=& \sin \frac{\theta}{2} \,
     \cosh\frac{\zeta}{2} \nonumber \\
   x &=& \sin \frac{\psi}{2} \, \sinh \frac{\zeta}{
     2} \nonumber \\
   y &=& \cos \frac{\psi}{2} \, \sinh \frac{\zeta}{
     2} \nonumber \\
   w &=& \cos \frac{\theta}{2} \, \cosh\frac{\zeta}{
     2}  \ . \eqnum{18}
\end{eqnarray}
The anti-de~Sitter metric on SU(1,1) is the metric induced by
Eq.~(16) on the submanifold defined by Eq.~(15). In the
coordinates $(\theta, \zeta,\psi)$, it is obtained by plugging
the transformations (18) into Eq.~(16), yielding
\FL
\begin{equation}
    ds^2=-\frac{1}{4}\cosh^2\frac{\zeta}{2}\ d\theta^2+\frac{1}{
     4}d\zeta^2 +\frac{1}{4}\sinh^2\frac{\zeta}{2}\
     d\psi^2 \ .\eqnum{19}
\end{equation}
Thus $\theta$ acts as a timelike coordinate.

\subsection{The Energy-Momentum Vector}

Elements of SU(1,1) sufficiently close to the identity may be
written as exponentials of elements of the Lie algebra:
\begin{equation}
    T=\exp{(-i\phi^\mu{{\cal J}}_\mu)} \ .\eqnum{20}
\end{equation}
Since $\phi^\mu$ describes a tangent vector at the identity of
SU(1,1), we can compute its norm using the metric defined above.
One could use Eq.~(19), but it is easier to expand $T$ to first
order in $\phi^\mu$, and then to compare with Eq.~(12) to
determine the parameters $(dt,dx,dy,dw)$ needed in Eq.~(16).  The
result is
\begin{equation}
    |\phi|^2 = \frac{1}{4} \eta_{\mu\nu} \, \phi^\mu \, \phi^\nu
     \ .  \eqnum{21}
\end{equation}
Not surprisingly, the group invariant metric in the vicinity of
the identity coincides, up to a factor, with the usual
(2+1)-dimensional Minkowski metric $\eta_{\mu\nu}$.  If
$|\phi|^2<0$ we will call the vector ``timelike,'' keeping in
mind the distinction between the timelike direction in the
spacetime manifold and that on SU(1,1). Note that $|\phi|$ is
actually the length of the curve defined by $T(\lambda)=\exp{(-i
\lambda \phi^\mu{{\cal J}}_\mu)}$, where $\lambda$ varies from 0
to 1, as can be seen by first calculating the length of the
segment from $\lambda$ to $\lambda + d \lambda$.

To understand the properties of $\phi^\mu$, let us write the
matrix describing parallel transport around a single particle in
the form $T=\exp(-i\phi^\mu{{\cal J}}_\mu)$. Under a Lorentz
transformation $L$ in the physical spacetime, the group element
$\exp(-i\phi^\mu{{\cal J}}_\mu)$ transforms as
\begin{equation}
    L\exp{(-i\phi^\mu{{\cal J}}_\mu)}L^{-1}=
  \exp(-i\Lambda^\mu{}_\nu\phi^\nu{{\cal J}}_\mu) \ ,\eqnum{22}
\end{equation}
where $L$ is an element of the $2\times 2$ representation of
SU(1,1) and $\Lambda$ is the corresponding matrix in the $3\times
3$ (adjoint) representation.  In the rest frame of a single
particle $T$ is a pure rotation and $\phi^\mu$ is equal to
$(\alpha,0,0)=(8\pi GM,0,0)$.  Using Eq.~(22), one sees that in
an arbitrary frame $\phi^\mu=8\pi G(\gamma M,\gamma M \vec v)$,
in agreement (up to a factor) with the energy-momentum vector of
special relativity.  It follows immediately that any massive
particle (moving slower than the speed of light) will be
associated with a $\phi^\mu$ that is timelike. If several
particles are combined, however, then $\phi^\mu_{tot}$ is not the
sum of the individual momenta; rather, from Eq.~(3),
\begin{eqnarray}
    \exp(-i\phi^\mu_{tot}{{\cal J}}_\mu)&=&\exp(-i\phi^\mu_N{{\cal
     J}}_\mu) \exp(-i\phi^\mu_{N-1}{{\cal J}}_\mu) \nonumber \\
    && \qquad \times \ldots \exp(-i\phi^\mu_1{{\cal J}}_\mu) \
     .\eqnum{23}
\end{eqnarray}
In the $G \to 0$ limit, on the other hand, each exponential can
be expanded to lowest order, and one finds that $\phi^\mu_{tot}$
approaches the sum of the individual $\phi^\mu_i$.  Since
$\phi^\mu_{tot}$ is a conserved Lorentz 3-vector which approaches
the ordinary special relativistic energy-momentum vector in the
$G \to 0$ limit, we will call it the energy-momentum vector.
However, it should be recognized that it is a somewhat
unconventional energy-momentum vector in at least two respects.
First, and more importantly for our purposes, the energy-momentum
vector for a group of particles is not the sum of the individual
momenta.  Second, it is constructed from the Lie algebra of the
Lorentz group, rather than the tangent space of the spacetime.
Nonetheless, in 2+1 dimensions the Lorentz generators can be
rearranged to form a vector, as in Eq.~(6).  The expansion of a
group element in terms of these generators gives rise to a
three-tuple $\phi^\mu$, which transforms according to Eq.~(22) as
a vector in the tangent space of the spacetime.  The tangent
space is constructed at the location of the base point of the
closed loop used for parallel transport, which might be thought
of as the location of the observer.\footnotemark \footnotetext{If
multiple observers are stretched along the loop, or along any
deformation of the loop that intersects no other particles, then
the observers will agree on the energy-momentum vector in the
following sense: if the vector measured by one observer is
parallel transported along the loop to another observer, it will
agree with the vector measured by the second observer.  We thank
Ken Olum for explaining this to us.}

For some purposes it will be convenient to write $\phi^\mu$ as a
parameter $\lambda$ times a normalized vector $n^\mu$; if
$\phi^\mu$ is timelike $|n|^2=-1$, and for $\phi^\mu$ spacelike
$|n|^2=+1$.  For $n^\mu$ timelike, the explicit form for $T$ is
\FL
\begin{eqnarray}
    e^{-i\lambda n^\mu{{\cal J}}_\mu}
     &=&\cos\frac{\lambda}{2}- 2in^\mu{{\cal
     J}}_\mu\sin\frac{\lambda}{2} \nonumber \\
   && \hskip -30pt = \left\lgroup \matrix{
     \cos\frac{\lambda}{2}-in^0\sin\frac{\lambda}{2}&
     (n^2+in^1) \sin\frac{\lambda}{2}\cr
    (n^2-in^1)\sin\frac{\lambda}{2}&
     \cos\frac{\lambda}{2}+in^0\sin\frac{\lambda}{2}\cr}
     \right\rgroup , \eqnum{24}
\end{eqnarray}
and for $n^\mu$ spacelike we obtain
\FL
\begin{eqnarray}
    e^{-i\lambda n^\mu{{\cal J}}_\mu}
     &=&\cosh\frac{\lambda}{2}- 2in^\mu{{\cal
     J}}_\mu\sinh\frac{\lambda}{2} \nonumber \\
   && \hskip -35pt = \left\lgroup \matrix{
     \cosh\frac{\lambda}{2}-in^0\sinh\frac{\lambda}{2}&
     (n^2+in^1) \sinh\frac{\lambda}{2}\cr
    (n^2-in^1) \sinh\frac{\lambda}{2}&
     \cosh\frac{\lambda}{2}+in^0\sinh\frac{\lambda}{2}\cr}
     \right\rgroup . \eqnum{25}
\end{eqnarray}
Taking the trace of these two equations, we can see that a
general matrix obtained by exponentiation satisfies
\FL
\begin{equation}
  \frac{1}{2}\mathop{\rm Tr}\nolimits \left[ e^{-i\lambda
     n^\mu{{\cal J}}_\mu}\right] =
  \cases{\cos\frac{\lambda}{2} & for $n^\mu$ timelike , \cr
  \cosh\frac{\lambda}{2} & for $n^\mu$ spacelike .\cr}\eqnum{26}
\end{equation}
It follows that
\begin{equation}
    \frac{1}{2} \mathop{\rm Tr}\nolimits \left[ e^{-i\lambda
     n^\mu{{\cal J}}_\mu}\right] \ge -1 \eqnum{27}
\end{equation}
for all cases.

{}From Eq.~(27), it is easy to show that the SU(1,1) matrix
corresponding to the Gott time machine is not the exponential of
any generator. For simplicity, we take a configuration where the
two particles approaching each other (as in Fig.~1) each have
rest frame deficit angle $\alpha$ and rapidity $\xi$, with
$\phi_1=\pi$, $\phi_2=0$.  Then we can use Eqs.~(3) and (9) to
write $T_G=T_2T_1$ as
\FL
\begin{equation}
    T_G = \left\lgroup \matrix{
     (1+p^2) e^{-i \alpha'} - p^2 & - 2 p \sqrt{1 + p^2} \sin
          \frac{\alpha'}{2} \cr
     - 2 p \sqrt{1 + p^2} \sin \frac{\alpha'}{2} & (1+p^2) e^{i
          \alpha'} - p^2 \cr} \right\rgroup , \eqnum{28}
\end{equation}
where $p$ and $\alpha'$ are given by Eqs.~(10) and (11).  The
trace is then given by
\begin{eqnarray}
    \frac{1}{2} \mathop{\rm Tr}\nolimits T_G &=& (1+p^2) \cos
     \alpha' - p^2 \nonumber \\
   &=& 1-2\cosh^2\xi \, \sin^2\frac{\alpha}{2}\ .\eqnum{29}
\end{eqnarray}
The condition that such a configuration contain closed timelike
curves is \cite{gott}
\begin{equation}
    \cosh\xi \, \sin\frac{\alpha}{2} >1 \ ,\eqnum{30}
\end{equation}
or $\frac{1}{2} \mathop{\rm Tr}\nolimits T_G<-1$.  Thus, from
Eq.~(27) it follows that $T_G$ cannot be written as an
exponential.

However, SU(1,1) is a double cover of the Lorentz group, so the
matrices $\pm T_G$ correspond to the same element of SO(2,1).  We
will see in the next section that $-T_G$ can be written as an
exponential.  Since $\frac{1}{2} \mathop{\rm Tr}\nolimits (-T_G)
>1$, Eq.~(26) implies that it is the exponential of a spacelike
generator.  The corresponding element of SO(2,1) can be obtained
by exponentiating the corresponding generator, and thus the
element of SO(2,1) associated with a Gott pair is spacelike or
tachyonic (equivalent under similarity transformation to a pure
boost).  This is the sense in which we say that the Gott time
machine has tachyonic momentum \cite{DJT,CFG}, even though
$T_G\in$ SU(1,1) is not equivalent to the exponential of a
spacelike generator --- parallel transport of a spinor around a
single tachyonic particle is not equivalent to parallel transport
around the Gott two-particle system, although parallel transport
of an SO(2,1) vector does not distinguish between the two cases.

\subsection{The Proof Completed}

We now return to the anti-de Sitter geometry of SU(1,1). For
fixed $n^\mu$, we may consider Eqs.~(24) and (25) as defining
curves parameterized by $\lambda$.  The crucial observation is
that these curves are geodesics in the metric (19), which can be
checked directly. For example, by comparing Eqs.~(17) and (24),
one sees that the curve defined by (24) is equivalent to
\begin{eqnarray}
    \theta&=&2\tan^{-1}\left(n^0\tan\frac{\lambda}{2}\right)
     \nonumber \\
    \zeta&=&2\sinh^{-1}\left[\sqrt{(n^0)^2-1}\sin\frac{\lambda}{2}
     \right] \nonumber \\
  \psi&=&-2\tan^{-1}\left({{n^1}\over{n^2}}\right)={\rm constant}\
     . \eqnum{31}
\end{eqnarray}
It is straightforward to confirm that this solves the geodesic
equation
\begin{equation}
    {{d^2x^\mu}\over{d\lambda^2}}+\Gamma^\mu_{\rho\sigma}{{dx^\rho}
  \over{d\lambda}}{{dx^\sigma}\over{d\lambda}}=0 \ ,\eqnum{32}
\end{equation}
for $x^\mu=(\theta,\zeta,\psi)$.  This fact can be seen more
directly by starting with a simple path, such as
$T(\lambda)=\exp(-i\lambda {{\cal J}}_0)$, and verifying that
this solves the geodesic equation.  Then by Eq.~(22) a Lorentz
transformation $\Lambda^\mu{}_\nu$ will take this curve into
another curve of the form $\exp(-i\lambda\phi^\mu {\cal J}_\mu)$,
with $\phi^\mu=\Lambda^\mu{}_0$.  Since the action of SU(1,1) is
an isometry, the resulting curve must also be a geodesic.
Finally, the isometry property also ensures that a curve of the
form $T(\lambda)= T_0\exp(-i\lambda n^\mu {\cal J}_\mu)$
($T_0\in$~SU(1,1)) will be a geodesic through $T_0$.

A simple way to visualize anti-de~Sitter space is in terms of its
Penrose (conformal) diagram \cite{hawkingellis}.  We define a new
coordinate $\zeta^\prime$ by
\begin{equation}
    \zeta^\prime=4\tan^{-1}\left(e^{\zeta/2}\right)-\pi \
     ,\eqnum{33}
\end{equation}
restricted to the range $0\leq\zeta^\prime<\pi$.  The metric (19)
becomes
\FL
\begin{equation}
    ds^2=\frac{1}{4 \cos^2 \frac{\zeta'}{2}} \left(-d\theta^2
     +{d\zeta^\prime}^2 + \sin^2\frac{{\zeta^\prime}}{2}\
     d\psi^2\right) \ .  \eqnum{34}
\end{equation}
The Penrose diagram is shown in Fig.~2; the angular coordinate
$\psi$ is suppressed. The light cones at each point are lines
drawn at $45^\circ$.  The right hand side of the rectangle is the
surface $\zeta^\prime=\pi$, which represents spacelike and null
infinity. The lower left corner is the origin, from which we have
drawn typical spacelike and timelike geodesics.  The lower and
upper boundaries are the surfaces $\theta=0$ and $\theta=4\pi$,
which are identified (the topology of SU(1,1) is thus $S^1\times
\relax{\rm I\kern-.18em R}^2$).  An important feature of this
diagram is that timelike geodesics from the origin refocus at the
point ($\theta=2\pi$, $\zeta^\prime =0$), as can be seen directly
from Eq.~(31) (note that $\zeta^\prime=0$ is equivalent to
$\zeta=0$).  Therefore, points that are spacelike separated from
($\theta=2\pi$, $\zeta^\prime =0$) cannot be joined to the origin
by a geodesic.  These points correspond to the shaded region of
the diagram.

Since every element of SU(1,1) that can be written as the
exponential of a generator lies along a geodesic through the
origin, points in the shaded region correspond to group elements
that cannot be reached by exponentiation.  The element $T_G$
corresponding to the Gott time machine lies in this region.  The
element $-T_G$, on the other hand, can be obtained from $T_G$ by
subtracting $2 \pi$ from $\theta$, so $-T_G$ lies in the region
that is spacelike separated from the origin.  Thus $-T_G$ can be
reached by exponentiation, as was claimed in the previous
section.

The product of two elements $T_B=\exp(-i\phi^\mu_B{{\cal
J}}_\mu)$ and $T_A=\exp(-i\phi^\mu_A{{\cal J}}_\mu)$ corresponds
to traveling in the direction of $\phi^\mu_A$ down a geodesic to
$T_A$, then traveling along a different geodesic (not through the
origin) to $T_BT_A$. As shown in the diagram, we can easily reach
the shaded region, and hence the Gott time machine, in this
manner.  The parameter space of SO(2,1) can be visualized by
cutting the diagram in half, identifying the surfaces $\theta=0$
and $\theta=2\pi$.  Then the shaded region of Fig.~2 is mapped
onto the wedge which is covered by spacelike geodesics emanating
from the origin.  This is consistent with our interpretation
above of the tachyonic momentum of the Gott time machine.

Consider a system of particles represented by an element $T_0$ of
SU(1,1).  We divide this system into a subsystem represented by
an element $T_S$ and the remaining $N$ individual particles
represented by $T_i$:
\begin{equation}
    T_0= T_N\ldots T_1 T_S \ .\eqnum{35}
\end{equation}
This relation can be represented on the Penrose diagram by a
future-directed nonspacelike curve from $T_S$ to $T_0$,
constructed from geodesic segments representing each of the
$T_i$. It is clear that the periodicity in the $\theta$ direction
allows any two points to be connected in this way --- as far as
SU(1,1) is concerned, any system of particles can contain a
subset with arbitrary energy and momentum.

However, the identification $\theta\leftrightarrow \theta+4\pi k$
obscures an important difference between physically distinct
situations.  To make this difference apparent, we must go to
$\widetilde{\hbox{SU}}\hbox{(1,1)}$, the universal cover of
SU(1,1).  $\widetilde{\hbox{SU}}\hbox{(1,1)}$ is then also the
universal cover of SO(2,1). In terms of the Penrose diagram, we
no longer identify $\theta=0$ with $\theta=4\pi$, but instead we
extend the picture infinitely far in the positive and negative
$\theta$ direction (Fig.~3).  The timelike geodesics from the
origin will refocus at $\theta=2\pi$, then continue onward,
refocusing again at $\theta=2\pi k$ for every integer $k$.
Therefore the wedges of points that are spacelike separated from
($\theta=2\pi k$, $\zeta^\prime=0$) for $k\neq 0$ cannot be
reached from any geodesic through the origin.  All of these
points may be said to correspond to tachyonic momenta, since they
map to elements of SO(2,1) lying on spacelike geodesics from the
origin.

To describe a multiparticle system using
$\widetilde{\hbox{SU}}\hbox{(1,1)}$, we express $T_{tot}$ as a
product as in Eq.~(3), with each $T_i$ representing a single
particle.  Any particle in its rest frame has a deficit angle
$\alpha < 2 \pi$, so one can uniquely define $T_i$ in the
universal covering group by using Eq.~(2), interpreting
$R(\alpha)$ as the element of the universal covering group
described by $(\theta = \alpha, \zeta' = 0)$, where $0 \le
\alpha < 2 \pi$.  Similarly $B(\vec \xi)$ can be chosen to lie in
the sector of the universal covering group that is spacelike
separated from the identity. (In this case, however, any other
choice would be equivalent.  The ambiguity consists of any number
of factors of the group element corresponding to a rotation by $2
\pi$, and this element commutes with all other elements.
Therefore, in Eq.~(2), the ambiguous factor would cancel between
$B$ and $B^{-1}$.)

Using this construction, a system of several particles, each with
timelike momentum, can lead to a $T_{tot} \in
\widetilde{\hbox{SU}}\hbox{(1,1)}$ in any region inside the
future light cone of the origin, including the wedges that are
spacelike separated from ($\theta=2\pi k$, $\zeta^\prime=0$) for
$k > 0$.  (A two-particle system, however, will only be able to
reach the first such wedge).

We now possess the machinery necessary to prove the theorem. We
consider an open universe, populated by particles with
future-directed timelike or lightlike momenta, such that the
total momentum is timelike. In such a universe we can always
boost into the rest frame, where (by definition) parallel
transport around all of the particles results in a rotation,
$R(\theta_0)$.  Since the universe is open, we must have
$\theta_0\leq 2\pi$.  Therefore, our hypothetical universe lies
on the straight line between the origin and ($\theta=2\pi$,
$\zeta^\prime=0$) in Fig.~3.  We now use the method of proof by
contradiction.  Suppose that some subset of particles in this
universe has a net spacelike momentum, corresponding to an
element $T_G$ of $\widetilde{\hbox{SU}}\hbox{(1,1)}$.  We can
then write
\begin{equation}
    R(\theta_0)=T_N\ldots T_1 T_G\ ,\eqnum{36}
\end{equation}
where the $T_i$ represent the individual particles comprising the
rest of the universe.  If we pick a point in the shaded region of
Fig.~3 to represent $T_G$, then the right hand side of this
expression must correspond to a point in the future light cone of
this point, since each $T_i$ can be represented by a segment of a
future-directed timelike or null geodesic.  However, it is clear
that none of the points on the straight line between the origin
and ($\theta=2\pi$, $\zeta^\prime=0$) lie in this light cone.
Therefore, we have a contradiction, and our supposition must be
false --- in an open universe with timelike total momentum, no
subset of particles can have a spacelike momentum.  This proves
the theorem; the statement in the title of our paper follows
immediately.

\section{DISCUSSION}

We have proven that a system of particles with spacelike momentum
can exist in an open (2+1)-dimensional universe only if the total
momentum is spacelike.  As the momentum of the Gott time machine
is necessarily spacelike, this result precludes the existence
(and hence the creation) of such a time machine in an open
universe with timelike total momentum.  In the process we have
found that the anti-de Sitter geometry of SU(1,1) provides a
useful tool for visualizing the momentum of a system of
gravitating particles in 2+1 dimensions.

There is an ongoing discourse concerning the appearance and
significance of CTC's in general relativity.  The causal
paradoxes which seem to inevitably accompany the existence of
time machines have led to the general belief that CTC's are to be
avoided --- although some recent investigations have argued to
the contrary \cite{friedmanetal}. If we adopt the conservative
attitude that CTC's lead to unacceptably paradoxical
consequences, we would hope that the laws of physics conspire to
prevent their occurrence.  While Einstein's equations permit a
wide variety of solutions that include CTC's, most of these
require either unrealistic energy-momentum tensors or global
identifications that appear to be physically unattainable (for
examples see Refs.~\cite{geroch} and \cite{hawkingellis}). We are
therefore led to ask whether CTC's can arise in an ``initially
causal'' universe.

Recent investigations of (3+1)-dimensional spacetimes with
traversable wormholes \cite{morrisetal} have shown that such
configurations seem to easily develop CTC's. However, the
maintenance of a traversable wormhole requires violation of the
weak energy condition (WEC).  While quantum field theory in
curved spacetime might allow WEC violation, there is evidence
that quantum fluctuations serve to destabilize the would-be time
machine, preventing the appearance of CTC's \cite{hawking,visser}
(for a competing view, see Ref.~\cite{kimthorne}).  Boulware
\cite{boulware} has studied the behavior of a scalar
(2+1)-dimensional quantum field theory in the Gott spacetime,
finding that if the mass of the quantum field is sufficiently
large, then the renormalized stress tensor is regular in the
vicinity of the Cauchy horizon. Given our incomplete
understanding of quantum gravity, it is worthwhile to ask if
classical general relativity (with the WEC enforced) can avoid
CTC creation.

Tipler \cite{tipler} has shown that creation of CTC's in a
compact region of a (3+1)-dimensional spacetime must necessarily
lead to the creation of a singularity, as long as the WEC holds
and there are tidal forces on some point on the Cauchy horizon.
(A similar theorem has been proven by Hawking in
Ref.~\cite{hawking}.) This theorem implies that an attempt to
make a time machine from finite lengths of cosmic string would
create a singularity, presumably (although not certainly)
surrounded by an event horizon (as suggested in
Ref.~\cite{gott}).  If the event horizon surrounds the CTC's,
they would not interfere with the causal properties of the
external space. The hypothesis that CTC's in 3+1 dimensions are
necessarily hidden behind horizons is reminiscent of the
connection between the ability to reach Gott's condition in 2+1
dimensions and the closure of the universe.

Tipler's and Hawking's use of the (Newman-Penrose) optical scalar
equations applies specifically to the (3+1)-dimensional case.
However, analogous reasoning can be adapted to 2+1 dimensions,
with the same result.\footnotemark \footnotetext{We are grateful
to Ted Pyne for discussions on this point.} Thus, this method can
be used to construct an alternative proof that Gott time machines
cannot be created in a local region of an open (2+1)-dimensional
universe.  This result, therefore, agrees with the theorem in
this paper and is significantly more general with regard to the
construction of time machines.

Meanwhile, the theorem proven in Sec.~II shows that a collection
of particles with a tachyonic momentum cannot even exist in an
open timelike universe, much less be created.  This theorem,
while limited to 2+1 dimensions, has no restriction analogous to
the compact-region requirement of Hawking's theorem, and it
applies to tachyons regardless of whether they are associated
with CTC's. Nevertheless, we have not excluded the possibility
that open timelike universes might contain CTC's distinct from
the type proposed by Gott. Waelbroeck \cite{waelbroeck} has shown
that a two-particle system with timelike momentum does not
support CTC's, and Kabat \cite{kabat} has presented arguments
suggesting that this result is more general.  A comprehensive
proof (or counterexample) is worth searching for.

Of course, Gott's original universe is still with us.  Does this
solution represent the creation of a time machine? It is
certainly not one we could build in the laboratory, as the CTC's
do not originate in a local region --- in the language of Hawking
\cite{hawking}, the Cauchy horizon is not compactly generated.
Nevertheless, there exist complete spacelike hypersurfaces which
are not intersected by CTC's, on which initial data can
presumably be specified \cite{cutler}.  Our work has shown that
we can place a restriction on this data such that a Gott time
machine can never arise. However, the restriction to timelike
momenta is rather intimately connected with the nature of
(2+1)-dimensional gravity, so it is not clear how this could be
promoted to a general principle applicable to the
(3+1)-dimensional case.

The demonstration that Gott time machines cannot exist in open
timelike (2+1)-dimensional universes leads naturally to the
question of closed universes.  In our earlier paper \cite{CFG} we
argued that the obstacle to building a time machine in an open
universe was that there could never be enough mass-energy to
accelerate particles to sufficiently high velocity; in other
words, it is impossible to produce two particles satisfying the
Gott condition by the decay of slowly-moving parent particles
unless the total rest frame deficit angle exceeds $2\pi$.  In a
closed universe, where the total deficit angle is $4\pi$, this
does not constitute an obstacle.  It is easy to imagine a closed
universe containing two particles, each with deficit angle
between $\pi$ and $2\pi$, and a number of less massive spectator
particles which bring the total deficit angle to $4\pi$.  Using
the description of decays given in Ref.~\cite{CFG}, we have found
that the two massive particles can decay in such a way that each
emits an offspring at sufficiently high velocity that the total
momentum of the two fast-moving particles is
tachyonic.\footnotemark \footnotetext{Our construction is
described in Ref.~\cite{thooft}.} Thus, in a closed
(2+1)-dimensional universe it is possible to ``build a tachyon.''
However, as we mentioned in the introduction, 't~Hooft
\cite{thooft} has shown that the size of the universe begins to
shrink precipitously after the decays, leading to a crunch (zero
volume) before any CTC's can arise.  There is thus a sense in
which general relativity is flexible enough to permit tachyons,
but works very hard to prevent time travel.

\nonum
\section{ACKNOWLEDGMENTS}

We thank Malcolm Anderson, Raoul Bott, Alex Lyons, Ken Olum, Don
Page, Ted Pyne, Gerard 't~Hooft, and David Vogan for very helpful
conversations. The work of S.M.C. was supported in part by funds
provided by the U.S. National Aeronautics and Space
Administration (NASA) under contracts NAGW-931 and NGT-50850, and
the work of E.F. and A.H.G. was supported in part by the U.S.
Department of Energy (D.O.E.) under contract \#DE-AC02-76ER03069.

\unletteredappendix{SOME PROPERTIES OF SU(1,1)}

In this Appendix we demonstrate two technical properties
concerning SU(1,1) and the embedding of its parameter space that
is introduced in \hbox{Sec.~II-C}.  First, we prove the statement
made in the text concerning the conditions under which a $2
\times 2$ matrix belongs to SU(1,1).  Next we demonstrate the
group invariance of the metric described in the text.

To derive the conditions under which a $2 \times 2$ matrix
belongs to SU(1,1), we begin by parameterizing an arbitrary $2
\times 2$ matrix as
\begin{equation}
    T = \left\lgroup \matrix{a & b \cr c & d \cr} \right\rgroup
     ,\eqnum{A.1}
\end{equation}
where $a$, $b$, $c$, and $d$ are all complex.  As stated in the
text, SU(1,1) consists of those matrices $T$ satisfying
\begin{equation}
    \det T=+1 \eqnum{A.2a}
\end{equation}
and
\begin{equation}
    T^\dagger \eta T= \eta \ ,\eqnum{A.2b}
\end{equation}
where
\begin{equation}
    \eta = \left\lgroup \matrix{1&0\cr 0&-1\cr} \right\rgroup .
     \eqnum{A.3}
\end{equation}
{}From Eqs.~(A.2) one has immediately that
\begin{eqnarray}
    a^*a - c^*c = 1 \eqnum{A.4a} \\
    b^*b - d^*d = -1 \eqnum{A.4b} \\
    b^*a - d^*c = 0  \eqnum{A.4c} \\
\noalign{\hbox{and}}
    ad - bc = 1 \ . \eqnum{A.5}
\end{eqnarray}

If Eq.~(A.4c) is solved for $c$ and the result is inserted into
(A.5), one finds
\[
    a (d^*d - b^*b) = d^* \ .
\]
Using (A.4b), this reduces to
\begin{equation}
    a = d^* \ .\eqnum{A.6}
\end{equation}
Combining this result with Eq.~(A.4c), one has
\begin{equation}
    b^* = c \ .\eqnum{A.7}
\end{equation}
Thus, $T$ can be written as
\begin{equation}
    T = \left\lgroup \matrix{a & b\cr b^* & a^*\cr} \right\rgroup
     , \eqnum{A.8}
\end{equation}
where from Eq.~(A.5) we have
\begin{equation}
    a^*a - b^*b = 1 \ .\eqnum{A.9}
\end{equation}
Comparing with the parameterization
\begin{equation}
    T = \left\lgroup \matrix{w - i t & y + i x\cr
             y - i x & w + i t\cr } \right\rgroup ,\eqnum{A.10}
\end{equation}
used in the text, one sees that $w$, $t$, $x$, and $y$ are all
real and satisfy
\begin{equation}
    -t^2 + x^2 + y^2 - w^2 = -1 \ . \eqnum{A.11}
\end{equation}

Conversely, it is easily shown that if $w$, $t$, $x$, and $y$ are
all real and satisfy Eq.~(A.11), then the matrix (A.10)
belongs to SU(1,1).

Next, we wish to verify that the metric defined in Sec.~II-C is
group invariant.  A group transformation on the group parameter
space can be defined by mapping each element of the parameter
space to the element obtained by multiplying on the left by a
fixed element of the group, which we call $\tilde T$.  Thus, the
mapping is defined by
\FL
\begin{equation}
    \left\lgroup \matrix{w' - i t' & y' + i x'\cr y' - i x' & w'
     + i t'\cr } \right\rgroup
     = \tilde T \left\lgroup \matrix{w - i t & y + i x\cr y - i x
     & w + i t\cr } \right\rgroup . \eqnum{A.12}
\end{equation}
Note that the metric of Eq.~(16) can be written as
\begin{eqnarray}
    ds^2 &=& -dt^2 + dx^2 + dy^2 - dw^2 \nonumber \\
     &=& \det (d T) \ , \eqnum{A.13}
\end{eqnarray}
where
\begin{equation}
    dT = \left\lgroup \matrix{dw - i dt & dy + i dx\cr dy - i dx
     & dw + i dt\cr } \right\rgroup . \eqnum{A.14}
\end{equation}
Since $\det \tilde T = 1$, it follows immediately that $\det (d
T') = \det (dT)$, so the metric is invariant.  It is similarly
clear that the metric is invariant under multiplication by a
fixed group element on the right.

(It is not needed in our derivation, but it is interesting to
note that the full invariance group of the metric given by
Eq.~(16) is SO(2,2), for which the Lie algebra is identical to
SU(1,1) $\times$ SU(1,1).  One of the two SU(1,1) subgroups has
generators that are self-dual (in the 4-dimensional
$w$-$t$-$x$-$y$ space), and the other has generators that are
anti-self-dual.  Transformations of the form described by
Eq.~(A.12) make up one of the SU(1,1) subgroups, while the other
subgroup corresponds to multiplication on the right.)

\newpage
\narrowtext

\figure{{\it A spacelike slice through the Gott spacetime.} Two
parallel cosmic strings perpendicular to the page, represented by
dots, move past each other at high velocity.  A deficit angle
(shaded) is removed from the space around each string, with
opposite sides identified at equal times.  (If the deficit angles
were oriented in any direction other than along the motion of the
string, the identifications would be at unequal times.) Note that
no CTC's pass through this spacelike surface, as explained in the
text.}

\figure{{\it The conformal diagram of SU(1,1), the double cover
of SO(2,1).} The group manifold of SU(1,1) with the invariant
metric is shown.  The $\psi$ direction is suppressed; hence, each
point away from the $\zeta^\prime=0$ line represents a circle.
The top and bottom edges, $\theta=4\pi$ and $\theta=0$, are
identified.  The identity element is in the lower left hand
corner; we have indicated some spacelike and timelike geodesics
from this point. Elements of SU(1,1) which can be expressed as
exponentials of generators lie on such geodesics.  The product
$T_G$ of two such elements $T_B$ and $T_A$ is represented by a
curve constructed from two consecutive geodesic segments, as
shown.  In this case the product lies in the shaded region, which
represents elements which cannot be expressed as exponentials of
generators.  The Gott time machine lies in this region.}

\figure{{\it The universal cover of SU(1,1).}  In the universal
covering group, the identification of $\theta$ with $\theta+4
\pi$ is removed, so that $\theta$ takes values from $-\infty$ to
$\infty$.  Thus, the interior of the future light cone of $T_G$,
which contains all universes consisting of the Gott time machine
and a set of other massive particles, does not include any points
on the $\zeta^\prime=0$ line between $\theta=0$ and
$\theta=2\pi$.  Since all open universes with timelike total
momentum lie on this line (in their rest frame), no such universe
can contain a Gott time machine.}

\end{document}